\documentclass[
superscriptaddress,
nofootinbib, twocolumn,
amsmath,amssymb,aps,
floatfix, eprint ]{revtex4-2}

\usepackage{graphicx}
\usepackage{dcolumn}
\usepackage{bm}
\usepackage{amsmath,amssymb,mathrsfs,amsfonts,slashed}
\usepackage{hyperref,xfrac}
\usepackage{url}
\usepackage{xcolor}
\newcommand{\p}{\bm{p}}

\def\tp{\tilde{p}}
\def\tq{\tilde{q}}

\def\tt{\tilde{t}}

\def\tpsi{\tilde{\psi}}
\def\mD{m_{\mathrm{D}}}
\def\mDz{m_{\mathrm{D},0}}

\begin{document}

\title{Medium-induced parton splitting in expanding QCD matter}

\author{Ismail Soudi}
\email{ismail.i.soudi@jyu.fi}
\affiliation{University of Jyväskylä, Department of Physics, P.O. Box 35, FI-40014 University of Jyväskylä, Finland.}
\affiliation{Helsinki Institute of Physics, P.O. Box 64, FI-00014 University of Helsinki, Finland.}

\date{\today}
\begin{abstract}
	Using the approach derived by Caron-Huot and Gale from the BDMPS-Z formalism, we obtain medium-induced parton splitting rates in a Bjorken expanding QCD matter.
	We compare the rate with the case of a static medium and investigate the impact of the medium expansion on the splitting rate.
	We also consider the leading order in the opacity expansion and the harmonic oscillator solution.
	These approximations are recovered in the limits of small and large formation times, respectively.
\end{abstract}

\maketitle
\section{Introduction}
One of the main signatures of quark-gluon plasma (QGP) is the suppression of high transverse momentum hadrons in heavy-ion collisions.
In addition to direct scattering with medium constituents, the parton shower is modified by medium-induced radiation.
Medium-induced radiation has been derived by Baier, Dokshitzer, Mueller, Peign\'e and Schiff (BDMPS) \cite{Baier:1994bd,Baier:1996kr,Baier:1996sk} and Zakharov (Z) \cite{Zakharov:1996fv,Zakharov:1998sv}, using the light-cone path integral formalism to resum the multiple-scattering diagrams that induce a splitting of the parton.

For a high-energy parton, soft splittings with a short formation time happen more quickly than the medium can resolve the individual quanta, making these splittings insensitive to the medium length.
In an infinitely large medium, the Arnold, Moore, and Yaffe (AMY) formalism \cite{Arnold:2008iy} was able to integrate out the time dependence of the BDMPS-Z formalism to obtain a simpler solution to the splitting rates.
These results were used extensively to understand energy loss by MonteCarlo simulations \cite{Schenke:2009gb} or effective kinetic descriptions \cite{Schlichting:2020lef,Mehtar-Tani:2022zwf}.

However, intermediate and hard splittings have a longer formation time, leading to multiple scatterings with the medium such that finite medium effects become important.
Caron-Huot and Gale introduced a formalism to investigate these finite-size effects on medium-induced splittings \cite{CaronHuot:2010bp}.
In this work, we adopt this framework to calculate the splitting rate within a Bjorken expanding medium.
Our approach is different from the one presented in \cite{Andres:2023jao}, which allows us to obtain a simpler numerical implementation.
This work is a foundation for future studies that will be able to utilize these rates to obtain a full resummation of the in-medium shower \cite{Schenke:2009gb,Kutak:2018dim,Mehtar-Tani:2018zba,Adhya:2019qse,Blanco:2020uzy,Schlichting:2020lef,Blanco:2021usa,Mehtar-Tani:2022zwf}.

Approximations to the in-medium splitting rate have been extensively explored in the literature, both in the context of static media \cite{Gyulassy:1999zd,Gyulassy:2000er,Mehtar-Tani:2019tvy,Mehtar-Tani:2019ygg,Barata:2020sav} and evolving media \cite{Bjorken:1982qr,Arnold:2008iy}.
In this work, we will introduce the opacity expansion (OE) and the harmonic oscillator (HO) approximations, which can be derived in a Bjorken expanding medium.
Our results will be compared with these approximations, and we will discuss their respective range of validity.

The paper is organized as follows:
The following section \ref{sec:Formalism} will introduce the formalism for medium-induced splittings in a static medium and extend it for the case of an evolving medium.
We also introduce the Opacity Expansion (OE) and the Harmonic Oscillator (HO) solutions to the evolution equation.
In section~\ref{sec:Results}, we will present the results for the splitting rate in an expanding medium and compare it to the static case.
We will also compare the full rate to the leading OE and HO solutions.
Finally, we will conclude with a summary of our important findings in Section~\ref{sec:Conclusions}.

\section{Medium-induced splitting functions in expanding QCD matter}\label{sec:Formalism}
Using the light-cone formalism, medium-induced splitting rates are obtained by solving a two-dimensional quantum evolution as derived in \cite{CaronHuot:2010bp} based on the BDMPS-Z framework \cite{Baier:1994bd,Zakharov:1996fv,Zakharov:1998sv}.
We consider the splitting of a quantum with energy $\omega = z P$ of a parton with energy $P$, in a medium with temperature $T$ and Debye screening mass $\mD = \frac{3}{2} g^2 T^2$ where $g$ is the QCD coupling.
Following earlier work \cite{Schlichting:2021idr} where the physical units were scaled out, the splitting rate for the process $a\to bc$ can be written as
\begin{align}\label{eq:RateEqInteraction}
    \frac{d\Gamma^a_{bc}}{dz}
    =&\frac{g^4 T P^a_{bc}(z)}{\pi } {\rm Re}~\int_0^{\tt} d{\Delta \tt}~ \int_{\tilde{\p}}  e^{-i\delta\tilde{E}(\tp) \Delta \tt} \tpsi_I(\tp,\Delta \tt) \;,
\end{align}
where we use the compact notation for the two-dimensional integral $\int_{\p}=\int \frac{d^2 p}{(2\pi)^2}$.
The interaction picture wave function $\tpsi_I(\tp,\Delta \tt)$ follows the evolution equation
\begin{align}\label{eq:EvolInteraction}
	\left[  \partial_{\Delta \tt} + \lambda\, e^{i\delta\tilde{E}(\tp) \Delta \tt}  \tilde{\p}\cdot \tilde{\Gamma}_3 \circ e^{-i\delta\tilde{E}(\tp) \Delta \tt} \frac{\tilde{\p}}{\tp^2} \right] \tpsi_I(\tp,\Delta \tt)  =&0\;,
\end{align}
with $\lambda =  \frac{2g^2TPz(1-z)}{\mD^2}$ counting the number of small angle scatterings per formation time of a splitting with small momentum transfer $\sim m_{D}$.
The initial condition of the wave function is given by
\begin{align}\label{eq:InitialInteraction}
    \tpsi_I(\tp,\Delta \tt=0) = \tilde{\p} \cdot \tilde{\Gamma}_3\circ  \frac{i\tilde{\p}}{\delta \tilde{E}(\tilde{\p})} \;,
\end{align}
where the convolution $\tilde{\Gamma}_3\circ$ describes the elastic interactions collision integral, written as,
\begin{align}\label{eq:ThreeBodyInt}
	&\tilde{\Gamma}_3(t) \circ G(t,\p) =\nonumber\\
	& \frac{1}{C_R}\int_{\tq} \Bigg[ C_a \tilde{C}(t,\tp - \tq) 
		+ \frac{C_b}{z^2}  \tilde{C}\left(t,\frac{\tp - \tq}{z}\right) \nonumber\\
	&+ \frac{C_c}{(1-z)^2}  \tilde{C}\left(t,\frac{\tp - \tq}{1-z}\right) \Bigg] \left[G(t,\tp) - G(t,\tq)\right]\;,
\end{align}
where $C_i$ the casimir factor of parton $i$ with $C_F=3$ and $C_A=8$ for quarks and gluons.
The energy of the quantum system is written
\begin{align}
	\delta \tilde{E}(\tp)
    =\tp^2 + (1-z)\frac{m_b^2}{\mD^2}+z\frac{m_c^2}{\mD^2} -z(1-z)\frac{m_a^2}{\mD^2}\;,
\end{align}
where $m_i$ are the thermal masses of the partons with $m_q^2 = C_F \frac{g^2 T^2}{4}$ and $m_g^2 = \frac{\mD^2}{2}$ for quarks and gluons respectively.
The tilde indicates that these quantities are dimensionless such that the dimensional quantities are given by
\begin{align}
	\Delta t =& \frac{2Pz(1-z)}{\mD^2} \Delta \tt\;,~~~~
	q = \mD \tq\;,~~~~
	p = \mD \tp\;,\\
	\delta E (\p) 
	=& \frac{\mD^2}{2Pz(1-z)} \delta \tilde{E}(\tp)\;.
\end{align}
The kernel $\tilde{C}(\tq)$ describes the transverse momentum broadening due to elastic scatterings, which in the leading order Hard Thermal Loop (HTL) approximation is given by \cite{CaronHuot:2008ni,Arnold:2008vd}
\begin{align}\label{eq:HTL}
	\tilde{C}(\tq)=
	\frac{\mD^2}{C_R g^2 T} C(q)
	=& \frac{1}{\tq^2(\tq^2 + 1)}\;.
\end{align}
Here we have taken a constant temperature $T$ to scale out the physical scales.
In the next section, we will reintroduce the temperature dependence of the medium to consider an evolving medium.

\subsection{Expanding QCD matter}
Let us consider a Bjorken expanding plasma where the temperature $T$ is related to the proper time $t$ by the following relation \cite{Bjorken:1982qr,Baier:1998yf}
\begin{align}
	T(t) = T_0 \left(\frac{t_0}{t + t_0}\right)^{\alpha/3}\;,
\end{align}
where the exponent $\alpha \leq 1$.
Although typically the denominator is taken to be $t$, we make explicit that the initial time $t_0$ is the entry point of the hard parton into the medium.
Our results will be presented such that for different initial times $t_0$, the temperature profile will be the same at the initial time $t=0$.
In order to make use of the dimensionless quantities from the previous section, we define a dimensionless temperature dependence $\tilde{T}(t) = \frac{T(t)}{T_0}$ in units of an initial temperature $T_0$.
The temperature and Debye mass used to scale out physical scales will be replaced by the constant quantities $T_0$ and $\mDz$.
The time dependence is obtained by inserting the proper powers of the dimensionless temperature $\tilde{T}(t)$.

The energy $\delta \tilde{E}(\tp)$ becomes time dependent and is given by 
\begin{align}
	&\delta \tilde{E}(\tt, \tp)\\
	&=\tp^2 + \left[(1-z)\frac{m_{z,0}^2}{\mDz^2}+z\frac{m_{1-z,0}^2}{\mDz^2} -z(1-z)\frac{m_{1,0}^2}{\mDz^2}\right] \tilde{T}^2(\tt)\;.\nonumber
\end{align}
Starting from the HTL kernel
\begin{align}
	\frac{1}{C_R}C(q)
	=& \frac{g^2T\mD^2}{q^2(q^2+\mD^2)}\;,
\end{align}
the time dependent elastic collision kernel $\tilde{C}(\tt,\p)$ can be written using the dimensionless kernel in Eq.~(\ref{eq:HTL})
\begin{align}
	\tilde{C}(\tt,\p)
	=\frac{\mDz^2}{C_Rg^2T_0}C(q)
	=&
	\frac{1}{\tilde{T}\left(\tt\right)}
	\tilde{C}\left(\frac{\tq}{\tilde{T}(\tt)}\right)\;.
\end{align}
Using this expression, the collision integral in Eq.~(\ref{eq:EvolInteraction}) becomes time dependent, evaluated at the complex conjugate amplitude time $\Delta \tt - \tt$, leading to the replacement
\begin{align}
	\Gamma_3 \to \Gamma_3(\Delta \tt - \tt)\;.
\end{align}
Conversely, the phase factor in the interaction term of Eq.~(\ref{eq:EvolInteraction}) will not be modified since the thermal masses cancel out between the two phase factors.

The initial wave function can no longer be written analytically as in Eq.~(\ref{eq:InitialInteraction}) and is instead given by the following integral \cite{CaronHuot:2010bp}
\begin{align}\label{eq:InitialInteractionExp}
	\tpsi_I(\tp,\Delta \tt=0; \tt) = \tilde{\p} \cdot \tilde{\Gamma}_3(\tt)\circ  \tilde{\p} \int_{\tt}^{\infty} ds~e^{-i\int_{\tt}^s du~ \delta \tilde{E}(u,\tp)}\;.
\end{align}
This integral can be resummed by rewriting it as a differential equation as described in Appendix~\ref{app:InitialCondition}.

The splitting rate in Eq.~(\ref{eq:RateEqInteraction}) becomes 
\begin{align}\label{eq:RateEqInteractionExp}
	\frac{d\Gamma^a_{bc}}{dz}
	=&\frac{g^4 T_0 P^a_{bc}(z)}{\pi } {\rm Re}~\int_0^{\tt} d{\Delta \tt}~ 
	\\
	&\int_{\tilde{\p}}  e^{-i\int^{\Delta \tt}_0 du~ \delta\tilde{E}(u,\tp)} \tpsi_I(\tp,\Delta \tt;\tt) \;,\nonumber
\end{align}
where the temperature prefactor stemming from the scaling of the physical units is now replaced by the initial temperature $T_0$.

The primary challenge in determining the rate compared to the static case arises from the time dependence of the initial condition described in Eq.~(\ref{eq:InitialInteractionExp}).
In contrast to the case of a static medium, where identical initial wave functions can be evolved and integrated at each time step to obtain the rate, in a dynamic medium it is necessary to establish a new initial condition for each time point and carry out the evolution separately.
The numerical implementation is described briefly in Appendix~\ref{app:InitialCondition}.

\subsection{Formation time}
Due to the uncertainty principle, the radiation quanta are emitted over a finite time given by the relation to the transverse energy \cite{Majumder:2010qh,Mehtar-Tani:2013pia}
\begin{align}
	t_f E_{\perp} \sim 1\;,
	\text{with}~E_{\perp} = \omega - k_z \simeq \frac{k_{\perp}^2}{2\omega}\;.
\end{align}
The formation time $t_f = \frac{2\omega}{k_{\perp}^2}$ and $k_{\perp}$ is the transverse momentum of the radiation.

Multiple elastic scatterings with the medium lead to transverse momentum broadening characterized by the transport coefficient $\hat{q}$ such that $\langle k_{\perp}^2\rangle \propto \hat{q}t$.
During the formation time, the transverse momentum of the radiation is then given by $k_{\perp}^2\simeq \hat{q} t_f$ leading to
\begin{align}
	t_f \simeq \sqrt{\frac{2zP}{\hat{q}}}\;.
\end{align}
Let us define the Bethe-Heitler scale $\omega_{BH} \equiv \frac{1}{2}\hat{q}\lambda^2_{\rm mfp} \simeq T$.
The interplay of the formation time and the mean free path $\lambda_{\rm mfp} \sim m_D^2/\hat{q}$ leads to two main regimes.
If the radiated energy $zP \ll \omega_{BH}$ or the time the hard parton spends in the medium $t\ll \lambda_{\rm mfp}$, the formation time is much smaller than the mean free path and the radiation is emitted before the parton can undergo multiple scatterings with the medium.
This is known as the Bethe-Heitler (BH) regime.

On the other hand for hard radiation $zP\gg \omega_{BH}$ with large formation time, the medium cannot resolve the individual quanta and multiple scatterings have to be resummed.
These multiple soft scatterings act coherently leading to a suppression of the radiation rate known as the Landau-Pomeranchuk-Migdal (LPM) effect \cite{Landau:1953um}.

\subsection{Leading order in opacity expansion}

In the BH regime, a finite number of scatterings occurs during the formation time such that the splitting rate can be obtained using an expansion in the number of the in-medium collisions.
This expansion is known as the opacity expansion (OE) or the Gyulassy, Levai, and Vitev (GLV) formalism \cite{Gyulassy:1999zd,Gyulassy:2000er}.
It is equivalent to a series expansion of the evolution equations in the number of collision operators.

Since we have absorbed the collision integral into the initial condition, the leading order term corresponding to a single scattering is readily obtained by plugging the initial wave function directly into the rate formula in Eq.~(\ref{eq:RateEqInteractionExp}), leading to
\begin{align}
	\frac{d\Gamma^a_{bc}}{dz}
	=&\frac{g^4 T_0 P^a_{bc}(z)}{\pi } {\rm Re}~\int_0^{\tt} d{\Delta \tt} \int_{\tilde{\p}}  e^{-i\int^{\Delta \tt}_0 du~ \delta\tilde{E}(u,\tp)}
	\nonumber\\
	& \tilde{\p} \cdot \tilde{\Gamma}_3(\tt)\circ  \tp \int_{\tt}^{\infty} ds~e^{-i\int_{\tt}^s du~ \delta \tilde{E}(u,\tp)} \;.
\end{align}
In contrast with the static case where the time integral can be performed analytically, here both time and momentum integrals will be performed numerically.
Similarly to the full rate, the initial condition in the opacity expansion is time dependent and must be computed for each time point before integrations.

\subsection{Deep LPM regime}

As the formation time is proportional to the energy of the radiation, the deep LPM regime is reached for large momentum fraction $z\sim 0.5$ and late times when both the formation time is large and the parton spends sufficient time in the medium to undergo multiple scatterings.

In this regime of a hard parton undergoing multiple scatterings, the elastic broadening kernel in Eq.~(\ref{eq:HTL}) is governed by small angle scatterings leading to transverse momentum broadening characterized by the transverse momentum diffusion coefficients \cite{Arnold:2008iy,Mehtar-Tani:2019tvy, Barata:2020sav}
\begin{align}
	\hat{\bar{q}}(t) = \frac{g^2 T(t) \mD^2(t)}{4\pi} \log \frac{aQ^2}{\mD^2}\;,
\end{align}
where $a=4e^{-2\gamma_E+1}\simeq3.42$, with $\gamma_E$ the Euler-Mascheroni constant.
We take half the geometric mean between the parton energy and the temperature of the medium $\left(Q^2=\frac{PT_0}{2}\right)$ as the scale for the ultraviolet cutoff.
The evolution equation in Eq.~(\ref{eq:EvolInteraction}) becomes a quantum harmonic oscillator (HO) problem with imaginary frequency $\omega(t)$ given by
\begin{align}
	\omega^2(t) =  -i\frac{[C_a + C_b z^2 + C_c(1-z)^2]\hat{q}(t)}{2Pz(1-z)}\;.
\end{align}

For a Bjorken expanding medium, where the solution to the rate equation was obtained in \cite{Bjorken:1982qr,Arnold:2008iy} for the spectrum
\begin{align}
	&\frac{dI^a_{bc}}{dz}
	= \frac{g^2}{4\pi^2 P} P^a_{bc}(z) \nonumber\\
	 &\ln \left| \left( \frac{t_0}{t_0 + t} \right)^{1/2}
	\frac{J_\nu(z_0)Y_{\nu-1}(z_t) - Y_{\nu}(z_0)J_{\nu-1}(z_t)}{J_\nu(z_t)Y_{\nu-1}(z_t) - Y_{\nu}(z_t)J_{\nu-1}(z_t)} \right|^2\;,
\end{align}
the rate can be recovered using a simple time derivate of the spectrum $\frac{dI^a_{bc}}{dz}$,
\begin{align}
	\frac{d\Gamma^a_{bc}}{dz}
	= \frac{d}{dt} \frac{dI^a_{bc}}{dz}\;.
\end{align}
Here $J_\nu$ and $Y_\nu$ are Bessel functions of the first and second kind and $\nu = \frac{1}{2-\alpha}$.

\subsection{Results}\label{sec:Results}
\begin{figure}
	\begin{center}
		\includegraphics[width=0.45\textwidth]{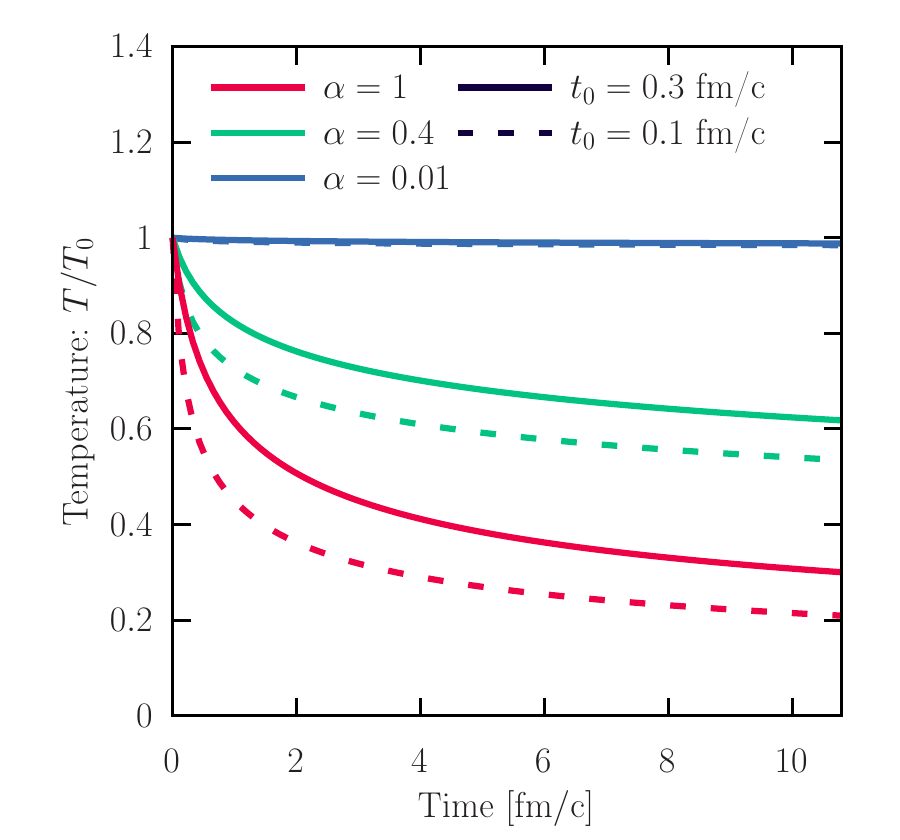}
	\end{center}
	\caption{
		Temperature profile for a Bjorken expanding medium with exponent $\alpha=1,0.4$ and $0.01$ represented by red, green and blue lines respectively.
		The initial time $t_0$ is chosen to be $t_0=0.3$ fm/c (full lines) and $t_0=0.1$ fm/c (dashed lines).
	}\label{fig:Temperature}
\end{figure}

The Bjorken expanding medium is characterized by the exponent $\alpha$ which determines the rate of expansion and the initial time $t_0$ when the hard partons enter the medium.
We present in Fig.~\ref{fig:Temperature} the temperature profile for different exponents $\alpha=1, 0.4$ and $0.01$ represented by red, green and blue lines respectively.
The initial time $t_0$ is chosen to be $t_0=0.3$ fm/c (full lines) and $t_0=0.1$ fm/c (dashed lines).
As expected, the exponential decay of the temperature is slowed down for smaller exponents $\alpha$ and larger initial times $t_0$.
When $\alpha=0.01$, the temperature remains virtually constant throughout the evolution.
This limit should correspond to the static medium case and we will use it as a cross-check for the numerical solver.

\begin{figure}
	\begin{center}
		\includegraphics[width=0.45\textwidth]{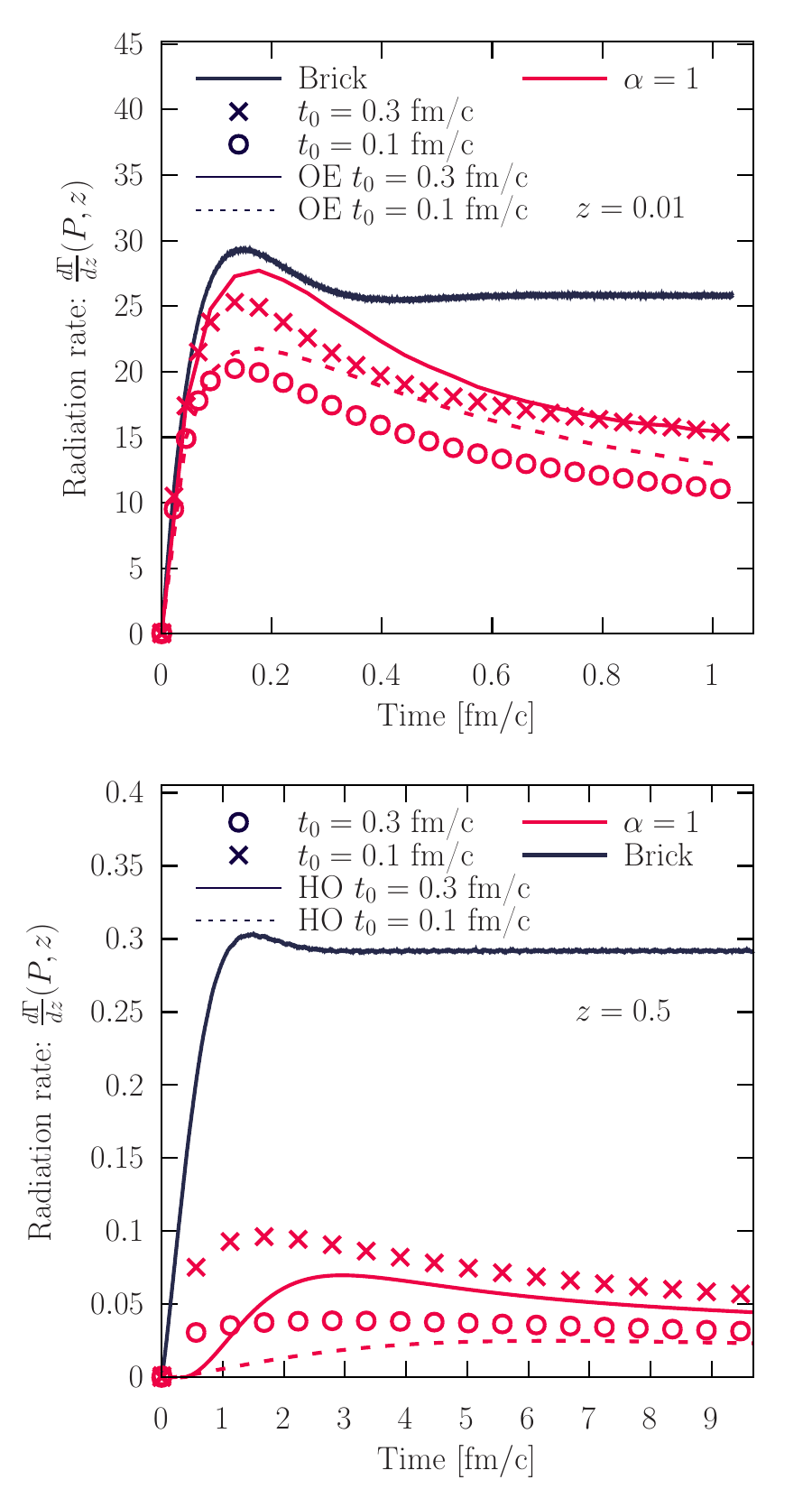}
	\end{center}
	\caption{
		Splitting rate of a gluon with momentum $P=16$ GeV to a gluon of momentum fraction $z=0.01$ (top) and $z=0.5$ (bottom).
		We compare the results in a static medium brick in black lines to a Bjorken expanding medium with exponent $\alpha=1$ and different initial time $t_0=0.3$~fm/c or $0.1$~fm/c using red circles or crosses respectively.
		In the top panel we compare with the first order OE in red dashed and full lines for the two initial times.
		In the bottom panel the HO solution is shown in red dashed and full lines for the two initial times.
	}\label{fig:InitialTime}
\end{figure}

\begin{figure}
	\begin{center}
		\includegraphics[width=0.45\textwidth]{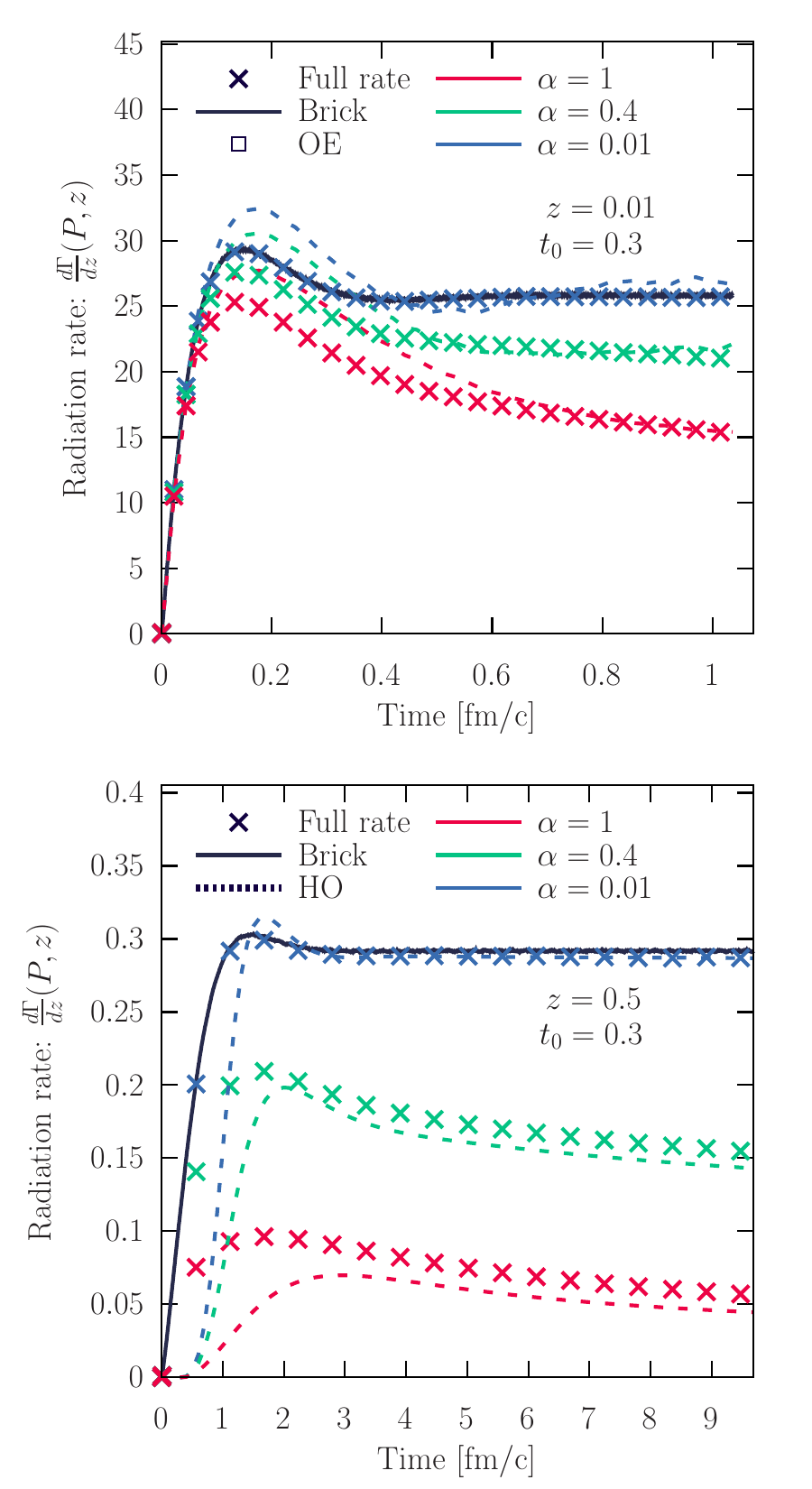}
	\end{center}
	\caption{
		Splitting rate of a gluon with momentum $P=16$ GeV to a gluon of momentum fraction $z=0.01$ (top) and $z=0.5$ (bottom).
		We compare the results in a static medium brick in black lines to a Bjorken expanding medium with exponent $\alpha=1,0.4$ and $0.01$ represented by red, green and blue crosses respectively.
		In the top panel we compare with the first order OE in dashed lines.
		In the bottom panel we compare with the HO solution in dashed lines.
	}
	\label{fig:Expansion}
\end{figure}

We consider the splitting of a gluon with momentum $P=16$ GeV to a gluon of momentum fraction $z=0.01$ and $z=0.5$.
We take the coupling constant $g=\sqrt{4\pi 0.3}\simeq 1.94$ and the temperature $T=0.4$ GeV.
The results will be compared to a static medium brick, to leading OE results and HO results.
We present two sets of figures, the first set in Fig.~\ref{fig:InitialTime} compares the splitting rate for different initial times $t_0=0.3$ fm/c and $0.1$ fm/c for a fixed exponents $\alpha=1$.
The second set in Fig.~\ref{fig:Expansion} compares the splitting rate for different exponents $\alpha=10^{-1},10^{-2}$ and $10^{-4}$ for a fixed initial time $t_0=0.3$ fm/c.

We start by investigating the impact of the initial time $t_0$ on the splitting rate.
For soft splitting $z=0.01$ in top panel of Fig.~\ref{fig:InitialTime}, we note that the initial linear growth is comparable between the static and Bjorken expanding medium for the full and OE rates.
However, the late time behavior is suppressed in the Bjorken expanding medium.
Conversely, the full rate recovers the OE results for $t_0=0.3$ fm/c, while for $t_0=0.1$ fm/c, the full rate remains below the OE results.
For small $t_0$ the temperature decreases rapidly leading to lower medium scales $\hat{q}$.
This causes the formation time ($t_f \propto \frac{1}{\sqrt{\hat{q}}}$) to increase leading a failure of the opacity expansion.

Conversely, for the hard splitting $z=0.5$ in the bottom panel of Fig.~\ref{fig:InitialTime}, the suppression of the splitting rate is more pronounced than for the soft splitting.
While the early-time behavior of the HO solution significantly differs from the full rate, the late-time behavior tends to converge towards the full rate. This convergence occurs because, at later times, the scale separation between the hard parton and the medium increases, resulting in a deep LPM regime.

In Fig.~\ref{fig:Expansion}, we note that the splitting rate with $\alpha=0.01$ aligns the static medium results for both the full rate and the approximate solutions in their respective range of validity.
Conversely, the expanding media with $\alpha=0.4$ and $1$ exhibits a suppression of the radiation rate compared to the static medium with this suppression being more pronounced for larger exponents and at late times.

For $z=0.01$ in the top panel, the opacity expansion is consistent with the full results at both early and late times.

For hard splitting, $z=0.5$ in the bottom panel, the difference between the exponents is larger leading to a strong suppression for $\alpha=1$.

\begin{figure}
	\begin{center}
		\includegraphics[width=0.45\textwidth]{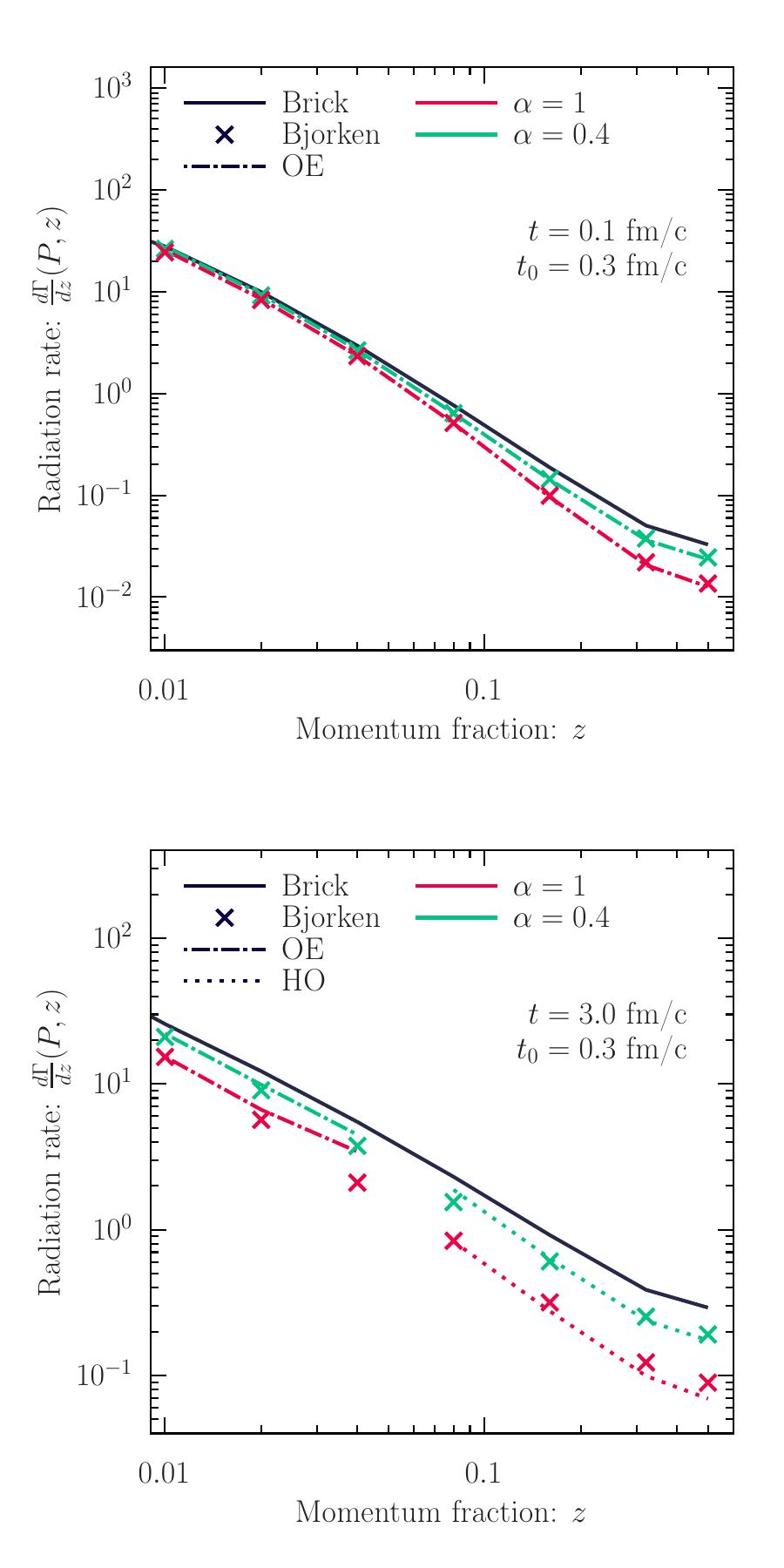}
	\end{center}
	\caption{
		Splitting rate as a function of the momentum fraction $z$ of the daughter gluon using $t_=0.3$ fm/c and $\alpha=1$ (red) and $\alpha=0.4$ (green).
		The top panel corresponds to an evolution time $t=0.1$ fm/c and the bottom panel to $t=0.3$ fm/c.
		The black lines correspond to the static medium brick, while the OE results are shown in dot-dashed lines and the HO results in dotted lines.
	}\label{fig:MomFract}
\end{figure}

Lastly, we present in Fig.~\ref{fig:MomFract} the splitting rate as a function of the momentum fraction $z$ of the daughter gluon.
We take $\alpha=1$ (red) or $0.4$ (green) and fix $t_0=0.3$ fm/c.
The two panels display two evolution times $t=0.1$ fm/c (top) and $t=0.3$ fm/c (bottom).

The results are consistent with the expectation that the opacity expansion is valid for small formation times or small momentum fractions.
We note that the suppression of the radiation rate is more pronounced for larger exponents $\alpha$ and at late time.
The deep LPM regime is reached at late times for momentum fractions $z\gtrsim 0.1$ and late times.

\section{Conclusions}\label{sec:Conclusions}
We have obtained medium-induced splitting rates in a Bjorken expanding QCD matter.
The results for a static brick medium are recovered in the limit of a small exponent $\alpha$ which serves as a cross-check for the numerical solver.
Conversely, we compared the full rate to the leading order in the opacity expansion and the harmonic oscillator solution, in their respective range of validity.
For early times or small momentum fractions, where the formation time is small, the OE results are consistent with the full rate.
Conversely, for late times and large momentum fractions, the HO solution is consistent with the full rate.

While we have focused on the splitting rate of a gluon to a gluon, all allowed QCD processes follow similar behavior and are obtained within this framework.
These results serve as a foundation for future studies that will be able to utilize these rates to obtain a full resummation of in-medium shower using either MonteCarlo approach or an effective kinetic description \cite{Mehtar-Tani:2022zwf}.

\section{Acknowledgments:}
 The author is indebted to K.~Eskola, O.~Garcia-Montero, A.~Majumder, Y.~Mehtar-Tani, H.~Niemi and S.~Schlichting for fruitful discussions throughout the evolution of this project.
 This work was funded as a part of the European Research Council project ERC-2018-ADG-835105 YoctoLHC, and as a part of the Center of Excellence in Quark Matter of the Academy of Finland (Projects No. 346325 and 364192).
 We acknowledge grants of computer capacity from the Finnish Grid and Cloud Infrastructure (persistent identifier urn:nbn:fi:research-infras-2016072533 ).

\appendix
\section{Numerical Implementation}\label{app:Numerical}
While the evolution of the wave function in a static medium can be carried out using various numerical schemes, the process becomes more intricate in an expanding medium. In this section, we will outline the numerical implementation of the wave function evolution in the expanding medium.

We first devise a quadrature rule to evaluate the momentum integrals.
The wave function will be evaluated at a set of discrete points $\tp_i$ such that any momentum integral can be approximated by a weighted sum over these points.

The angular integral of Eq.~(\ref{eq:RateEqInteraction}) can be evaluated analytically, leading to the following expression \cite{Andres:2020vxs}
\begin{align}
	&\int_0^{2\pi} d\theta~\tilde{C}_{\rm QCD}\left(t,\frac{\tp - \tq}{z}\right)\\
	&=  \frac{1}{|\tp - \tq|} - \frac{1}{\sqrt{(z^2+\tp^2+\tq^2)^2 - 4\tp^2\tq^2}}
	\;,\nonumber\\
	&\int_0^{2\pi} d\theta~ \cos\theta ~\tilde{C}_{\rm QCD}\left(t,\frac{\tp - \tq}{z}\right)\\
	&=\frac{1}{2\tp\tq} \left[\frac{\tp^2+\tq^2}{|\tp - \tq|} - \frac{z^2+\tp^2+\tq^2}{\sqrt{(z^2+\tp^2+\tq^2)^2 - 4\tp^2\tq^2}}\right]
	\;.\nonumber
\end{align}
The collision integral becomes a simple $\tq$ integral that can be evaluated by a weighted sum of the wave function at each point.

The time integrals present in the phase factors can be evaluated analytically to obtain
\begin{align}
	&\int_{0}^{\tt} du~\delta \tilde{E}(u,\tp)
	= \tp^2 u +
	M_{\rm eff} t_0^{\frac{2\alpha}{3}}
	\\
	&\times
	\frac{
	\left(\tt + t_0\right)^{\frac{2\alpha}{3}} \left(\tt + t_0 - \Delta \tt\right)
	-\left(\tt + t_0\right) \left(\tt + t_0 - \Delta \tt\right)^{\frac{2\alpha}{3}}
	}{\left[\left(\tt + t_0\right)\left(\tt + t_0 - \Delta \tt\right)\right]^{\frac{2\alpha}{3}}\left(\frac{2\alpha}{3}-1\right)}\;.\nonumber
\end{align}

The wave function is evolved using a simple Euler's scheme with a time step $d\Delta \tt$.
We then integrate the wave function at each time point to obtain the rate.

\subsection{Initial condition}\label{app:InitialCondition}
The integral in Eq.~(\ref{eq:InitialInteractionExp}) can be difficult to evaluate directly.
Instead, one can rewrite it as a differential equation by differentiating with respect to $\tt$.
We start by defining the function
\begin{align}
	F(\tt,\tp)
	=& \int_{\tt}^{\infty} ds~e^{-i\int_{\tt}^s du~ \delta \tilde{E}(u,\tp)}\;.
\end{align}
The time derivative of this function is given by
\begin{align}
	\partial_{\tt} F(\tt,\tp)
	=& -1 + i\delta \tilde{E}(\tt,\tp)\int_{\tt}^{\infty} ds~e^{-i\int_{\tt}^s du~ \delta \tilde{E}(u,\tp)}\;,\\
	=& -1 + i\delta \tilde{E}(\tt,\tp)F(\tt,\tp)\;.
\end{align}
We employ a change of variables $x = 1/\tt$, with $\frac{d}{d\tt} = -x^2\frac{d}{dx}$, to rewrite the differential equation as
\begin{align}
	x^2\partial_{x} F(x,\tp)
	=& 1 - i\delta \tilde{E}(1/x,\tp)F(x,\tp)\;.
\end{align}
Using an inverse Euler's scheme with inverse time step $dx$, the following solution is obtained
\begin{align}
	F(x + dx, \tp)
	= \frac{\frac{F(x,\tp)}{dx} + \frac{1}{(x+dx)^2}}{\frac{1}{dx} + \frac{i\delta \tilde{E}\left(\frac{1}{x+dx},\tp\right)}{(x+dx)^2}}\;.
\end{align}
By simply taking the initial condition $F(x=0,\tp) = 0$, one can ensure the integral is convergent at $\tt\to\infty$.
The function $F(x,\tp)$ is then obtained by iterating this equation from $x=0$ to the time when the rate is evaluated at $x=1/\tt$.

\bibliography{main.bib}
\end{document}